\documentstyle[11pt]{article}
\renewcommand {\topfraction}{0.8}
\renewcommand {\textfraction}{0.2}
\renewcommand {\floatpagefraction}{0.7}
\newtheorem{theorem}{Theorem}
\newtheorem{lemma}{Lemma}
\def\appendix{\par				
    \setcounter{section}{0}			
    \setcounter{subsection}{0}
    \def\thesection{Appendix \Alph{section}}
    \renewcommand{\theequation}{\Alph{section}.\arabic{equation}}
}

\begin{document}

\title{Exact Results for the Asymmetric Simple Exclusion Process with a
	Blockage\thanks{
                Supported in part by NSF Grant DMR92-13424}}
\author{S. A. Janowsky\thanks{Supported in part by NSF Mathematical
                Sciences Postdoctoral Research Fellowship DMS
                90-07206}\hspace{.25em}\thanks{Address after August 1993:
		Department of Mathematics, University of Texas,
		Austin, TX 78712}\hspace{.25em}
        and J. L. Lebowitz\\[.3cm]
        Departments of Physics and Mathematics\\
        Rutgers University\\
        New Brunswick, New Jersey 08903}
\date{June 1993}
\maketitle
\renewcommand{\baselinestretch}{1.5}\large\normalsize
\vspace{-.5cm}
\begin{abstract}
We present new results for the current as a function of transmission
rate in the one dimensional totally asymmetric simple exclusion process
(TASEP) with a blockage that lowers the jump rate at one site from one
to $r<1$.  Exact finite volume results serve to bound the allowed values
for the current in the infinite system.  This proves the existence of a
gap in allowed density corresponding to a nonequilibrium ``phase
transition'' in the infinite system.  A series expansion in $r$, derived
from the finite systems, is proven to be asymptotic for all sufficiently
large systems.  Pad\'e approximants based on this series, which make
specific assumptions about the nature of the singularity at $r=1$, match
numerical data for the ``infinite'' system to a part in $10^4$.

\end{abstract}
\renewcommand{\baselinestretch}{1.7}\large\normalsize

\section{Introduction}
The one dimensional totally asymmetric simple exclusion process (TASEP)
is a continuous-time sto\-chastic process in which particles on a one
dimensional lattice jump independently and randomly at unit rate to
vacant neighboring sites on their immediate right~\cite{reviews}.  It
corresponds to a Kawasaki exchange dynamics~\cite{Kawa} at infinite
temperature and infinite electric field~\cite{KLS}.  The stationary
state of this system for $N$ particles on a ring of $K$ sites, $K\ge N$,
gives equal weight to all $K\choose N$ permissible configurations.  This
measure goes over, in the limit $K\rightarrow\infty$, $N/K \rightarrow
\rho$, to the product measure with occupation probability $\rho$.  The
TASEP is thus the simplest driven diffusive lattice-gas model whose
dynamics does not satisfy detailed balance~\cite{KLS}.  It is also, for
the infinite lattice, an example of a microscopic system from which one
can derive Euler-like hydrodynamical equations~\cite{dMP}, {\em e.g.}\
the Burgers equation.

In an earlier work~\cite{JL} we introduced a variant of the TASEP where
the jump rate across one bond of the system was reduced from 1 to $r$,
$0<r<1$.  If one thinks of the TASEP as a model for fluid flow in a
pipe, this is analogous to a restriction in the diameter of the pipe.
More realistically perhaps we could consider a superionic
conductor~\cite{DFP} like AgI or KAg$_4$I$_5$, in the geometry of a
pinched doughnut, with a time-varying magnetic field generating an
electromotive force; or a road under construction in a model of traffic
flow~\cite{NS92}.  Clearly the rate decrease will increase the particle
density to the immediate left of this ``blockage'' bond and decrease the
density to its immediate right, but what is not obvious is that this
perturbation may have global in addition to local effects.  Equivalent
to introducing a defect into a growth surface~\cite{WT,KM}, this
blockage can cause the nonequilibrium stationary states of the model to
exhibit a segregation into high- and low-density regions.  This allows
the full complexity of the model, previously available only through
time-dependent studies, to be displayed via the stationary
state~\cite{JL,ACJL}.  Among other features we observed for the system
of $N$ particles on a ring of size $K$, with $N$ and $K$ large, was the
equivalence of the exponent governing the time and space scaling of
shock fluctuations.

\begin{figure}
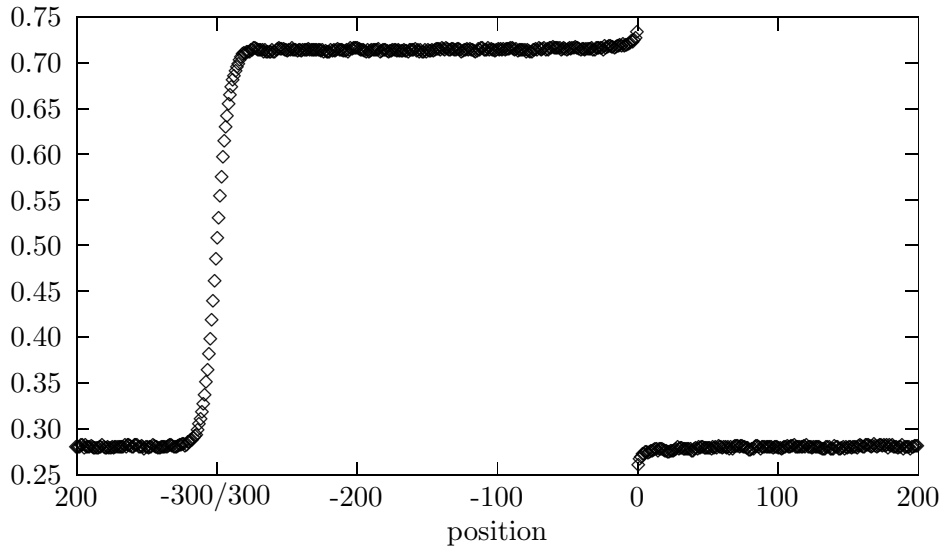

\begin{center}
\setlength{\unitlength}{0.2409pt}
\ifx\plotpoint\undefined\newsavebox{\plotpoint}\fi
\sbox{\plotpoint}{\rule[-.175pt]{.35pt}{.35pt}}%

\end{center}
\caption{Density profile (from time average in simulation) for a
half-filled system with 600 sites and periodic boundary conditions.  The
blockage bond is located between site 0 and site 1 and has the value
$r=0.33$.}
\label{profile}\end{figure}

It is convenient to label the sites on the ring from $-K/2$ to $K/2$
($-K/2$ and $K/2$ refer to the same site), with the blockage
located at the bond between 0 and 1.  The stationary density profile for
a half-filled system can be seen in fig.~\ref{profile}.
As we approach the limit $K\rightarrow\infty$ keeping $N/K = \rho$
fixed, we have that far from the blockage the density becomes uniform
and the stationary state appears to be asymptotically a product
measure~\cite{Bramson} with densities $\lim_{x\rightarrow\pm\infty}
\rho(x) = \rho_\pm$, with $\rho_+ < \rho_-$, independent of $\rho$, for
$|\rho - 1/2|$ less than a certain value described in the next section.  The
densities are related by the condition of constant current: $J =
\rho_+(1-\rho_+) = \rho_-(1-\rho_-)$.  When segregation takes place we
thus have $1/2 - \rho_+ = \rho_- - 1/2$.  One is of course interested in
the asymptotic densities $\rho_\pm$ and current $J$ as a function of the
transmission rate $r$.  In \cite{JL} these could only be determined
numerically; the simple estimate obtained by neglecting the (large)
correlations near the blockage, {\em i.e.}\ by assuming a uniform
product measure with density $\rho_-$ (resp.\ $\rho_+$) to the left
(resp.\ right) of the origin, gives
\begin{equation}\label{badapprox}
J \approx r \rho_- (1-\rho_+) \approx \frac{r}{(1+r)^2},
\end{equation}
which is accurate only for very small $r$.  This is in contrast with the
translation invariant case, where, as already mentioned, there are no
correlations in the stationary state.  Here, in order to obtain more
accurate estimates, it is necessary to include the effects of
correlations, which decay slowly with the distance from the blockage,
typical of systems with conservative dynamics not satisfying detailed
balance~\cite{BS,GLMS,Grinstein}.  (In \cite{JL} this decay was found
numerically to go like the inverse power of the distance.)

Although one is interested in asymptotic behavior, it can be fruitful to
examine the behavior of small systems with open boundary conditions for
which one can determine the stationary state exactly.  In a closely
related model~\cite{DDM}, this led to an exact solution for all system
sizes.  Furthermore, as we shall see, results for such systems give a
systematic improvement on the estimate (\ref{badapprox}) as well as
bounds on the current $J$ for the infinite system at fixed $r$ and thus
also on the maximum value of $\rho_+$.  We therefore studied systems,
ranging in size from 2 to 10 sites, in which particles are added at the
left with rate $\alpha$ and removed at the right with rate $\beta$.  At
non-boundary sites particles jump independently to empty neighboring
sites on the right with rate $r$ for the jump between sites 0 and 1,
where the blockage is located, and with rate $1$ for all other jumps.
The process is thus defined (on a lattice with $K=2L$ sites) by the
generator $\cal L$ giving the rate of change of any function of the
configuration $\eta = \{\eta(-L+1), \eta(-L+2),
\ldots, \eta(L)\}$, where $\eta(k) = 0\ {\rm or}\ 1$ is the occupation
number at site $k$:
\begin{eqnarray}
{\cal L} f(\eta) &=& \alpha [f(\eta_{-L+1,+}) - f(\eta)]
        + \sum_{i=-L+1}^{-1}[f(\eta_{i\rightarrow i+1}) - f(\eta)]
                \nonumber\\
        && + r[f(\eta_{0\rightarrow 1}) - f(\eta)]
        + \sum_{i=1}^{L-1}[f(\eta_{i\rightarrow i+1}) - f(\eta)]
                \label{generator}\\
        && + \beta [f(\eta_{L,-}) - f(\eta)],\nonumber
\end{eqnarray}
where $\eta_{i\rightarrow j}(k)$ gives the configuration at site $k$
after an attempted jump from site $i$ to site $j$:
\begin{equation}
\eta_{i\rightarrow j}(k) = \left\{
\begin{array}{l@{\quad}l}
1,& k=j {\rm ~and~} \eta(i) = 1,\\
0,& k=i {\rm ~and~} \eta(j) = 0,\\
\eta(k),& {\rm otherwise};
\end{array}
\right.\end{equation}
the boundary terms are given in terms of
\begin{equation}
\eta_{i,\pm}(k) = \left\{
\begin{array}{l@{\quad}l}
(1\pm1)/2, & k =  i,\\
\eta(k),   & k\ne i.
\end{array}
\right.\end{equation}

\section{Exact Solutions For Small Systems}
For a system with $2L$ sites there are $2^{2L}$ possible configurations.
By considering the set of equations $\langle{\cal L} f_i\rangle = 0$
where $f_i$ is the characteristic function of the $i$th configuration,
one obtains a system of equations whose (normalized) solution is the
unique stationary state for the model, {\em i.e.}\ one obtains the
unique stationary solution of the master equation for the open finite
system.

Solving such a system of equations is a daunting computational task.
However, by fixing $\alpha$ and $\beta$ and then making use of the
reflection symmetry of the problem (which for $\alpha=\beta$ reduces the
number of variables by a factor slightly less than 2) we were able to
exactly solve for systems up to size 8 {\em leaving $r$ as an
indeterminate parameter} and size 10 for fixed $r$.  The computations
were performed using Maple V running on a SPARCstation 2.

Typically we took $\alpha =\beta = 1$; the asymptotic behavior should be
independent of $\alpha$ and $\beta$ provided they are greater than some
critical value which goes to zero as $r$ goes to zero and equals 1/2 for
$r\ge1$.  This critical value is simply related to the maximum current
the infinite system with blockage $r$ can support; it is just the
density of a product measure at the maximum current: $(1-\sqrt{1-4J_{\rm
max}})/2$.  Clearly $J_{\rm max}(r)$ is a monotone nondecreasing
function of $r$ equal to $1/4$ for $r\ge 1$.  The system ``selects'' the
state of maximum current, $\rho_+ = (1-\sqrt{1-4J_{\rm max}})/2$ and
$\rho_- = 1-\rho_+$, if the boundaries can supply and remove particles
quickly enough~\cite{Krug}, so the (asymptotic) state should not depend
on the precise values of $\alpha$ and $\beta$; corrections to this
asymptotic behavior will be local to the boundary
region~\cite{DDM,DEHP}.  Alternatively, if one considered periodic
boundary conditions instead of an open system we would expect equivalent
asymptotics with densities $\rho_- = 1-\rho_+$ for all values of the
average density $\rho$ between $(1-\sqrt{1-4J_{\rm max}})/2$ and
$(1+\sqrt{1-4J_{\rm max}})/2$.  For average densities outside this
interval the limiting asymptotic densities should be equal to the
average density both to the left and right of the blockage (see
fig.~\ref{phase_r-fig}).

We present some of the results for small systems below.  $J_L(r,a)$ is the
steady state current in a system with $2L$ sites at $\alpha = \beta =
a$; $J_L(r) \equiv J_L(r,1)$.  We have
\begin{equation}\label{J1-J1a}
J_1(r,a) = \frac{2ar}{2a + 3r}, \qquad
J_1(r) = \frac{2r}{2 + 3r}, \qquad
J_1(r,1/2) = \frac{r}{1 + 3r};
\end{equation}
\begin{eqnarray}
J_2(r,a) &=& 2ar \left[ 4a^2 + 8a^3 + 4a^4 + \left(
        7a + 12a^2 + 6a^3 \right)r + \left(2 + 4a + 2a^2\right)r^2
                \right]         \Bigg/  \nonumber\\[2pt]
&&      \Big[ 8a^3 + 16a^4 + 8a^5 + \left( 14a^2 + 36a^3 + 36a^4 + 12a^5
        \right)r        \\
&&\quad  + \left(18a + 45a^2 + 44a^3 + 18a^4 \right)r^2 +
        \left( 5 + 14a + 14a^2 + 6a^3 \right)r^3        \Big],\nonumber
\end{eqnarray}
\begin{equation}
J_2(r) = \frac{2r\left(16 + 25r + 8r^2\right)}{
         32 + 98 r + 125 r^2 + 39 r^3}, \qquad
J_2(r,1/2) = \frac{2r (9 + 29 r + 18 r^2)}{18 + 85 r + 215 r^2 + 130 r^3};
\end{equation}
\begin{eqnarray}
J_3(r) &=& 4r\big(47775744 + 261095424 r + 669424384 r^2 +
                1009680576 r^3 + 968982368 r^4
                        \nonumber\\
&&\quad  + 609395274 r^5 + 250834237 r^6 + 65287925 r^7 + 9784215 r^8
        + 644781 r^9    \big) \Big/     \nonumber\\[2pt]
&&      \big( 191102976 + 1331036160 r + 4447316992 r^2 + 9277942272 r^3
\label{J3}\\
&&\quad         + 12731145304 r^4 + 11671707972 r^5 + 7170513506 r^6
                + 2914237861 r^7        \nonumber\\
&&\quad          + 753023405 r^8 + 112354075 r^9 + 7383541 r^{10}
                        \big).  \nonumber
\end{eqnarray}
We observe that $J_L(r)$ is a rational function of $r$ with integer
coefficients, and the order of the function and the complexity of the
coefficients grow rapidly with system size---$J_4(r)$ has terms up to
$34$th order with the largest coefficient being 51 digits long, and is
reproduced in \ref{J4}.  We note that the value of $J_L(1)$ was obtained
explicitly in \cite{DDM} for all $L$, namely that
\begin{equation}\label{DDM-result}
J_L(1)	= \frac14 + \frac3{4(1+4L)}
	= J_\infty(1) \left[ 1+\frac3{1+4L} \right].
\end{equation}

\section{Bounds On The Infinite System}
For $r=1$, product measure with any density $\rho\in [0,1]$ is
stationary in the infinite system, and thus the system can have any
current in the range $0$ to $1/4$. For $r<1$, this is not necessarily
the case; $J_{\rm max}(r)$ may be less than $1/4$.  In this case only a
range of densities satisfying $|\rho_\pm - 1/2| > \sqrt{1-4J_{\rm
max}(r)}/2$ is permitted.  The problem thus is to find bounds on $J_{\rm
max}(r)$.  A very simple bound on $J_{\rm max}(r)$ can be obtained by
noting that if we remove the right (or left) half of the system we are
left with a system of $L$ sites with input (removal) rate $a$ and
removal (input) rate $r$.  Calling $\hat{J}(L; \alpha,\beta)$ the
current in a system of $L$ sites with input rate $\alpha$, removal
rate $\beta$, and all ``internal'' jump rates now being unity,  we
clearly have
\begin{equation}
J_L(r,a) \le \hat{J}(L; a,r) = \hat{J}(L; r,a),
	\label{halfsys}
\end{equation}
where the last equality, as well as exact formulae for $\hat{J}(L;
\alpha,\beta)$, have been computed in \cite{DDM,DEHP}.  In particular
for $a\ge r$ and $r\le 1/2$ we have $\lim_{L\rightarrow\infty}\hat{J}(L;
\alpha,r) = r(1-r)$ so that $J_{\rm max}(r) \le r(1-r)$ for $r\le
1/2$.

For $a$ and $r$ both greater than $1/2$ the right hand side of
(\ref{halfsys}) approaches $1/4$ so it yields no new information---just
a proof that $J_{\rm max}(r)$ cannot exceed $1/4$.

\begin{figure}
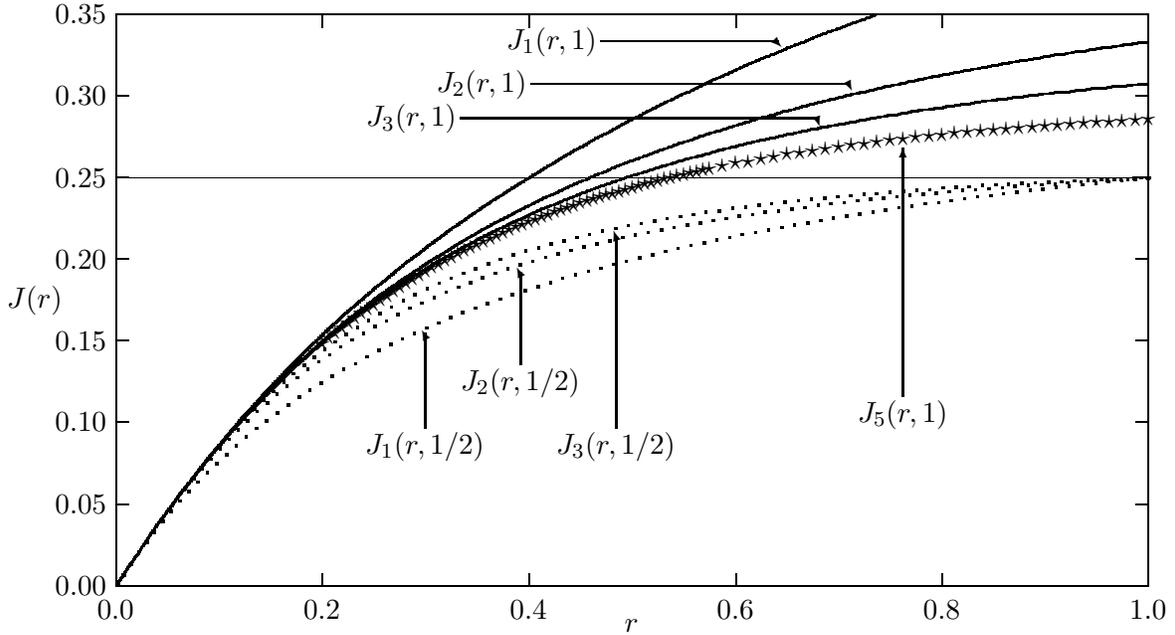

\begin{center}
\setlength{\unitlength}{0.2409pt}
\ifx\plotpoint\undefined\newsavebox{\plotpoint}\fi

\end{center}
\caption{$J_L(r,1)$ and $J_L(r,1/2)$ for several values of $L$.  Where
$J_L(r,1) < 1/4$ there must be a gap in the allowed stationary measures
of the infinite system.}
\label{finite}\end{figure}

To obtain better bounds we note that for any configuration, since the
maximum rate at which a particle attempts to jump on to any site is
bounded by one, the current in a system of size $L$ with boundary
conditions $\alpha=\beta=1$ cannot increase as $L$ increases.  Note that
$\alpha=\beta=1$ corresponds to keeping the site $-L$ always occupied
and the site $L+1$ always vacant in a system of size $L'>L$.  Thus
$J_L(r)$ is monotonically decreasing in $L$, and for every finite value
of $L$, $J_L(r)$ is an upper bound for $J_\infty(r)$.  (Similar
arguments show that $J_L(r,a)$ is nondecreasing in $r$ and $a$.)  These
bounds prove the existence of a gap in the set of stationary measures of
the infinite system; the results for finite $L$ (illustrated in
fig.~\ref{finite}) show that wherever there exists an $L$ such that
$J_L(r) < 1/4$ there is a forbidden range of currents.  (One result from
our simulations is that $J_{15000}(0.8) = 0.24979(5)$, indicating that
the gap exists at least up to $r=0.8$.)

We note here also that letting $a\rightarrow\infty$ in $J_L(r,a)$
corresponds to reducing the size of the system from $2L$ to $2(L-1)$ so
that $\lim_{a\rightarrow\infty} J_L(r,a) = J_{L-1}(r).$  Using
monotonicity in $a$ this implies again that $J_L(r)$ is monotone
nonincreasing in $L$.

For the case $\alpha=\beta=1/2$ the sequence $J_L(r,1/2)$ appears to be
increasing in $L$, and we believe but cannot prove that $J_\infty(r)\ge
J_L(r,1/2)$.  Proving such an inequality is nontrivial since
$J_\infty(r)$ is the {\em maximum} current the infinite system can have;
there exist stationary measures where the current is less than
$J_L(r,1/2)$.  Thus our results for finite $L$ only provide rigorous
one-sided bounds.  Note also that $J_L(1,1/2) = 1/4$ for all $L$,
c.f.~\cite{DDM}.

\subsection*{Series expansion}

While there is no apparent pattern to the ``raw'' expressions for the
current (\ref{J1-J1a})--(\ref{J3}), (\ref{J4-eq}), one
does emerge if we examine a Taylor expansion around $r=0$.  We obtain:
\begin{eqnarray}
\label{J1Taylor}
J_1(r) &=& r - \frac32 r^2 + \frac94 r^3 - \frac{27}8 r^4 +
        \frac{81}{16} r^5 - \frac{243}{32} r^6 + O(r^7),        \\[5pt]
J_2(r) &=& r - \frac32 r^2 + \frac{19}{16} r^3 - \frac{257}{256} r^4 +
        \frac{24105}{4096} r^5 - \frac{829297}{65536} r^6 + O(r^7),\\[5pt]
J_3(r) &=& r - \frac32 r^2 + \frac{19}{16} r^3 - \frac{21535}{27648} r^4 +
        \frac{919407829}{214990848} r^5 -
        \frac{7398899579671}{417942208512} r^6 + O(r^7),        \\[5pt]
J_4(r) &=& r - \frac32 r^2 + \frac{19}{16} r^3 - \frac{21535}{27648} r^4 +
        \frac{77729356627}{146767085568} r^5	\nonumber\\[-6pt]
\label{J4Taylor}\\[-6pt]
&&\quad + \frac{1067903077191004635349}{126214320739011526656} r^6
        + O(r^7).\nonumber
\end{eqnarray}
There is a clear pattern: as we increase the size of the system the
low-order coefficients stop changing after a certain point.  Assuming
the continuation of this behavior, and
including the results for the size 10 system, we have
\begin{eqnarray}
J_\infty(r)\!
&=& r - \frac32 r^2 + \frac{19}{16} r^3 - \frac{21535}{27648} r^4 +
        \frac{77729356627}{146767085568} r^5 -.3278724755(1) r^6 + O(r^7)
        \nonumber\\[-8pt]
\label{taylor}\\[-4pt]
&=&  r - \frac32 r^2 + \frac{19}{2^4} r^3 -
                \frac{5\cdot 59\cdot73}{2^{10}\cdot 3^3} r^4 +
                \frac{13\cdot 33613\cdot 177883}{2^{26}\cdot 3^7} r^5
                 -.3278724755(1) r^6 + O(r^7);\nonumber
\end{eqnarray}
in the second part of (\ref{taylor}) we show the prime factorization of
the coefficients; the denominators appear deceptively simple while
examination of the numerators proves less instructive.

Including the dependence of the boundary terms ({\em i.e.}\ taking $a\ne
1$) does not significantly alter this behavior:
\begin{eqnarray}
J_1(r,a) &=& r - \frac3{2a} r^2 + \frac9{4a^2} r^3 -\frac{27}{8 a^3} r^4
	+ O(r^5),\\
J_2(r,a) &=& r - \frac32 r^2 + \frac{9 a^4  + 18 a^3  + 7 a^2 -8 a - 7}{
	4 a^2 (a+1)^2} r^3\nonumber\\
&&{}- \frac{
	54 a^7 + 216 a^6 + 300 a^5 + 64 a^4 - 294 a^3 - 370 a^2 - 188 a -39}{
	16 a^3 (a + 1)^4} r^4 + O(r^5),\\
J_3(r,a) &=& r - \frac32 r^2 + \frac{19}{16} r^3 + \Big[514 a^{14}
	 + 6939 a^{13} + 41551 a^{12} + 144387 a^{11} +
	316671 a^{10}\nonumber\\
&&{} + 432661 a^9 + 285181 a^8 - 176743 a^7 - 702157 a^6 - 944908 a^5
 - 799104 a^4 - 457504 a^3\nonumber\\
&&{} - 172480 a^2 - 38528 a - 3840\Big]r^4 \Big/
	\left[256 a^3 (1 + 2a)(a+2)^3 (a+1)^7\right] + O(r^5).\label{J3aT}
\end{eqnarray}
The dependence on the boundary appears one term earlier than in
(\ref{J1Taylor})--(\ref{J4Taylor}), but otherwise the structure is the same.

In fact, the assumption regarding the behavior of the Taylor
coefficients for progressively larger systems expressed in
(\ref{J1Taylor})--(\ref{J3aT}) can be proven to be correct:
\begin{theorem}\label{asymtheorem}
Fix $L_2$.  Then for $L_1 \le L_2$, $J_{L_2}(r) = J_{L_1}(r) + O(r^{L_1 + 2})$.
\end{theorem}
\begin{theorem}\label{asymtheorem2}
Fix $L_2$.  Then for $L_1 \le L_2$ and $a>0$,
$J_{L_2}(r,a) = J_{L_1}(r) + O(r^{L_1 + 1})$.
\end{theorem}
Thus we see a small system not only bounds but also provides a good
approximation (at least for small $r$) for any (finite) system that is
larger.  This also strongly suggests that the approach of $J_L(r)$ to
$J_\infty(r)$ is exponential for $r<1$.  Of course the rate of the
exponential approach vanishes at $r=1$, where the convergence becomes
algebraic (see (\ref{DDM-result})).
We reserve the proof for \ref{proof}.

\section{Pad\'e Approximants}
The Taylor series given in (\ref{taylor}) gives an accurate measure of
the current for small $r$, but for large $r$ it is less successful.  In
fact, we can see that the series must break down by $r=1$: although one
commonly thinks of $r$ as being a transmission rate $\le 1$, it is
perfectly acceptable to take $r>1$ in the generator (\ref{generator}).
It is also fairly easy to see, by comparison with a system with a
boundary at the origin, that the infinite volume current $J_\infty(r)$
will have the same value for all $r\ge 1$, namely $J_\infty(1) = 1/4$,
so that there must be a nonanalyticity for some $r\le 1$ in
$J_\infty(r)$.  One can also examine the coefficients of (\ref{taylor})
and see that they apparently decrease (in magnitude) rather slowly,
indicating that the radius of convergence of the series is most probably
1.  Some numerical analysis indicates that there is no discontinuity in
any of the derivatives of $J_\infty(r)$ at $r=1$; this evidence leads us
to hypothesize that there is an essential singularity at $r=1$ in
$J_\infty(r)$.  An alternative which cannot be ruled out by our results
is that $J_\infty(r) = 1/4$ for $r>r_1$ with $r_1<1$.  Numerical results
for the structure of $J_L(r)$ would seem however to argue in favor of a
changeover at $r=1$.

We thus look for a function of the appropriate form for the current.
The ``simplest'' function with an essential singularity at $r=1$ that
also gives $J_\infty(0) = 0$ and $J_\infty(1) = 1/4$
is a function of the form
\begin{equation}
J_\infty(r) = 1/4 - \exp[f(r)]/4,
\end{equation}
where $f(r)$ has a simple pole at $r=1$.  We therefore examined
functions of the form
\begin{equation}
f(r) = \frac14 - \frac14\exp\left[\frac{p(r)}{(1-r)q(r)}\right];
\end{equation}
specifically, we used our Taylor series to fit the following Pad\'e
approximants:
\begin{eqnarray}
J^{33}(r) &=& \frac14 - \frac14\exp\left[\frac{-4r(1 + a_1 r + a_2 r^2)}{
        (1-r)(1 + b_1 r + b_2 r^2)}\right],     \\[5pt]
J^{43}(r) &=& \frac14 - \frac14\exp\left[\frac{-4r(1 + a_1' r + a_2' r^2 +
        a_3' r^3)}{(1-r)(1 + b_1' r + b_2' r^2)}\right],        \\[5pt]
J^{34}(r) &=& \frac14 - \frac14\exp\left[\frac{-4r(1 + a_1'' r + a_2'' r^2)}{
        (1-r)(1 + b_1'' r + b_2'' r^2 + b_3'' r^3)}\right].
\end{eqnarray}

The actual values of the coefficients of the Pad\'e approximants are not
particularly revealing, as one might expect.   The Pad\'e functions do,
however, appear to be converging pointwise:  for $r\in [0,1]$ the
maximum difference between any of $J^{33}$, $J^{43}$ and $J^{34}$ is
less than $2\times 10^{-5}$.

\begin{figure}
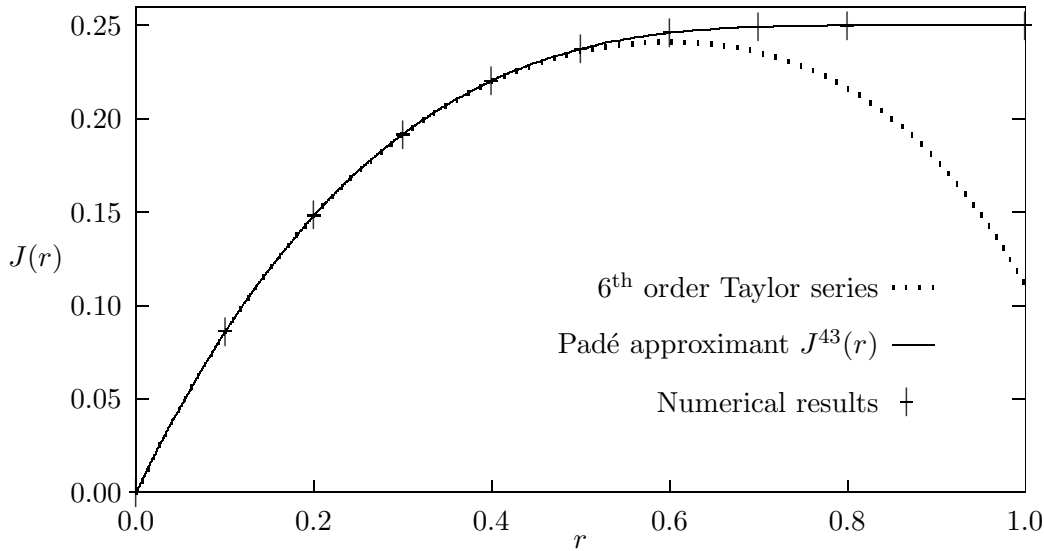

\begin{center}
\setlength{\unitlength}{0.2409pt}
\ifx\plotpoint\undefined\newsavebox{\plotpoint}\fi
\sbox{\plotpoint}{\rule[-.175pt]{.35pt}{.35pt}}%

\end{center}
\caption{Taylor, Pad\'e and numerical results.  Estimated errors for the
numerical calculations are less than the thickness of the lines, ranging
from $5\times 10^{-5}$ to $2\times 10^{-4}$.}
\label{J-r-fig}
\end{figure}

Of course we are not interested in how well the different approximants
approximate each other, but how well they approximate $J_\infty(r)$.  We
thus must compare the approximants with numerical simulations.  We see
in fig.~\ref{J-r-fig} that the approximation is within the error bounds
of the simulations; we happened to have plotted $J^{43}$ but any of the
approximants would have fit the data as well.  In comparison, a Pad\'e
approximant that behaves quadratically or quartically at $r=1$ (as
opposed to exponentially) looks qualitatively similar but does not fall
within the error bounds of the simulations.

The simulation results were obtained by direct simulation of the TASEP
dynamics; progressively larger systems were used until the current
reached an asymptotic value.  For small values of $r$ this occurred
quite quickly, but the needed system size grows quite rapidly as one
approaches $r=1$: for $r=0.7$ we needed to investigate systems with 6400
sites and for $r=0.8$ we needed to investigate systems with 15000 sites.

Our results can also be viewed as the determination of the phase diagram
for the system.  As mentioned earlier the current $J_\infty(r)$ is the
{\em maximum} stationary current permitted by the infinite system---with
some boundary conditions the current may be less.  Accepting
\cite{Bramson} that the stationary measure is asymptotically a product
measure then $J_\infty = \rho_\infty(1-\rho_\infty)$, and the bound on
the current is equivalent to a bound on the allowable range of
densities.  We plot the boundary of this range, the critical density,
against the transmission rate
\begin{figure}
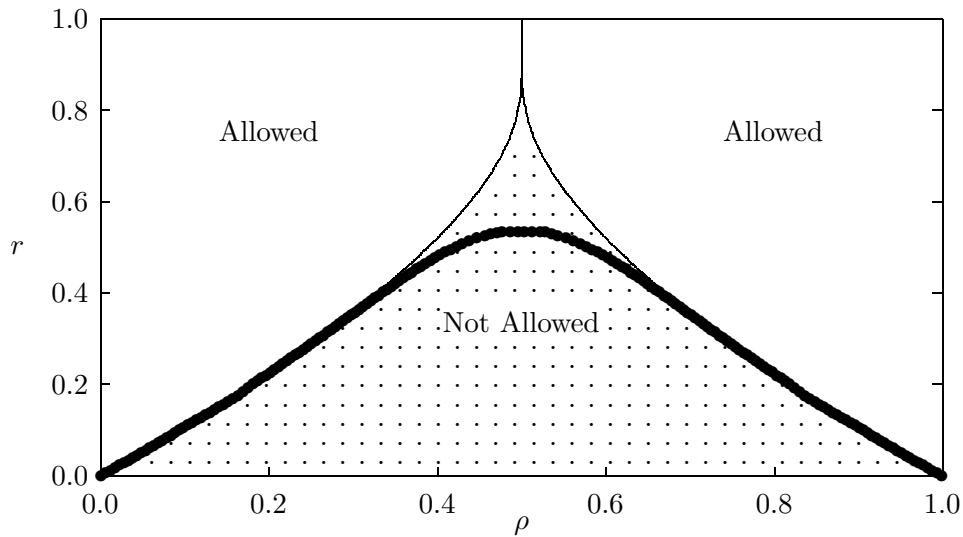

\begin{center}
\setlength{\unitlength}{0.2409pt}
\ifx\plotpoint\undefined\newsavebox{\plotpoint}\fi
\sbox{\plotpoint}{\rule[-.175pt]{.35pt}{.35pt}}%

\end{center}
\caption{Phase diagram of allowed asymptotic density {\em vs.}\
transmission.  The heavy line indicates the region excluded by currently
available exact results; the thin line and the shaded region are
determined by the Pad\'e  results.}
\label{phase_r-fig}
\end{figure}
in fig.~\ref{phase_r-fig}.  If the overall density of the finite system
in a periodic box, $\rho$, was in the disallowed region, the system
would have to segregate with high density $\rho_-$ to the left of the
blockage and low density $\rho_+$ to the right of it, with $\rho_- +
\rho_+ = 1$ and $\rho = c\rho_- + (1-c)\rho_+$ for some fraction $c$,
$0<c<1$.

The phase diagram is apparently quite different from that normally
observed in systems that phase segregate, with a cusp at the critical
point if one believes the Pad\'e approximants accurately describe the
behavior of the system.  Of course one should remember the rather unique
nature of this type of transition.

\section{Discussion}
When a system has dynamics that do not satisfy detailed balance, the
steady states do not (in general) have the form of Gibbs states.  Thus,
even a qualitative description of the nature of the steady states,
particularly with respect to the dependence on the parameters entering
the dynamics, is lacking.  Standard perturbative techniques are generally
inapplicable, and most of our knowledge comes from computer simulations,
approximate calculations using renormalization group (universality)
ideas, and a few exactly soluble models.

It thus is quite remarkable that information determined from a system
with 10 sites can be used to predict the behavior of a system with
$10^4$ sites to within a part in $10^4$ for all values of the parameter
$r$.  This is particularly interesting given that quantities besides the
asymptotic current, {\em e.g.}\ the local density near the blockage, are
not given accurately by the results from the small systems.

We hope that it will be possible to determine a general formula for the
terms in our Taylor series---certainly the sequence of denominators
($1,\, 2,\, 2^4,\, 2^{10} 3^3,\, 2^{26} 3^7$) seems tractable.  Since
the coefficients are independent of system size (if the system is big
enough) only reasonably-sized systems need to be studied in order to
obtain useful results.  Unfortunately direct computation of higher-order
terms seems unlikely, given that the computational complexity grows
exponentially, without further theoretical input.

Even if our computations cannot be extended, there remains the
possibility of proving that some of our qualitative description of the
phase diagram is correct.  Our series is only proven to be asymptotic
for small $r$, but we have already seen that finite systems can serve to
bound certain properties of the infinite system for arbitrary values of
$r$: for example we believe that $J_L(r,1/2) \le J_\infty(r)$ and we
know that $J_\infty(r)\le J_L(r,1)$ for all $L>0$ and all $0\le r\le 1$.

\renewcommand{\baselinestretch}{1}\large\normalsize
\appendix
\renewcommand{\thesection}{Appendix \Alph{section}}
\section{The function $J_4(r)$}\label{J4}
\setcounter{equation}{0}
$J_4(r) =$
\begin{equation}\label{J4-eq}
{\scriptsize\frac{
\begin{array}{c}
2r\cdot\big(
46594570041553208865114157983928672r^{34} +\\
3272256807527488050658050196010194008r^{33} +
111157036726242688664509666570575952692r^{32} +\\
2432978692972890221457116915305608070176r^{31} +
38563042893543905939312199308038000267376r^{30} +\\
471629643887135562049943024025386831761624r^{29} +
4630830690747494363914241555175156474737952r^{28} +\\
37501116089445029526599189649856279114367080r^{27} +
255349882255809833270215616484973179447971604r^{26} +\\
1483034668343761760933311314169905866593307200r^{25} +
7427024657245470907468866632968858025038621224r^{24} +\\
32341852979577601689520720487367994974417590384r^{23} +
123258959443962771434544300470844688653283339220r^{22} +\\
413187952609538688738716350764927640900234526696r^{21} +
1222925851525408590841791369809880777445578325556r^{20} +\\
3204670630666608783584495910901402180619797197008r^{19} +
7449424997833123151978707539741892038741323482132r^{18} +\\
15378015964501993765574252081802020182175729230264r^{17} +
28201950376870887559972172287807374530071235163796r^{16} +\\
45931932106228251497346831830246712730250907133712r^{15} +
66367346077080103767105589115636872559989214550441r^{14} +\\
84922961375392487277902239726735636329302249943128r^{13} +
95985607018161363007960499193162341535912061049744r^{12} +\\
95495899491963130911426581603975014896548923261952r^{11} +
83250539026363674821927230131182759997615987109888r^{10} +\\
63222693774680812665162035869098746973122020442112r^9 +
41512233585213949665760166256158119415119899262976r^8 +\\
23337512260240138424809534978398421177693582131200r^7 +
11088988610809857322221895274024151043294063755264r^6 +\\
4375523126050494396079936095098710196366170128384r^5 +
1398264149827785557648870482515232428810049683456r^4 +\\
348505991819536772424594468421800136094590697472r^3 +
63703892226452562097992685657738901012902576128r^2 +\\
7610495543799775598845584755474871742038540288r +
446443113770610009364511999835819193606864896\big)
\end{array}}
{
\begin{array}{c}
\big(
289620280319036422192318811402048808r^{35} +
20347660058076078580887803589027266100r^{34} +\\
691509180090958636822780801273385528578r^{33} +
15143171306946285250897143672457580856112r^{32} +\\
240156757626348422508288967223318773219730r^{31} +
2939013682631225236236144632768990651743996r^{30} +\\
28878453872089870194234547560112273497410248r^{29} +
234055024231518053113128876309994685465865998r^{28} +\\
1595219951014207692709829729475814422350403348r^{27} +
9274911880978847804107170124884025274187759308r^{26} +\\
46507371429345382752005370961862741048375464600r^{25} +
202819905961978964692531821554688508829966150639r^{24} +\\
774305913910573093515211395294773846677490470023r^{23} +
2600902978686057892600556891176773816130767065580r^{22} +\\
7716564464804046735459659032377460737774938262294r^{21} +
20279584392798156542425998004346657157480554084969r^{20} +\\
47304820202361946730446972991019456869802234988211r^{19} +
98064294514560130382369885216651184747849062158254r^{18} +\\
180769182668553783438231100925433451981380541595764r^{17} +
296287842239143809420984903211913226411657347513603r^{16} +\\
431494541664969696255270976652324836280276021938501r^{15} +
557613386760147331351017000992052900309321526028898r^{14} +\\
638162101110332154357279278430725658354907392715680r^{13} +
645076256680960920649709526072469900202092833324704r^{12} +\\
573956167631320085801939332326120871725603509536768r^{11} +
447552460656613237448046131746142189157548579971072r^{10} +\\
304176946559301647309430508519196663546007393796096r^9 +
178937028617243446184563625330908654841920346914816r^8 +\\
90290026293392171865547954006547530297965379846144r^7 +
38609332952461801047936627883941955240573198663680r^6 +\\
13757984465655294150768749140417907410320746348544r^5 +
3986654552595495753325700874754842100218571259904r^4 +\\
904824015385555468140765882234506884944960159744r^3 +
151187962701066997262421713581553533038335361024r^2 +\\
16560320428911381225784705510457201064897675264r +
892886227541220018729023999671638387213729792\big)
\end{array}}}
\end{equation}

\renewcommand{\baselinestretch}{1.7}\large\normalsize
\section{Proof of Theorem~\protect\ref{asymtheorem}}\label{proof}
\setcounter{equation}{0}
To prove theorem~\ref{asymtheorem}, we will need to consider systems
with unequal numbers of sites to the left and to the right of the
blockage, say $L_-$ and $L_+$, respectively; $J_{L_-,L_+}(r)$ will
represent the current in such a system with $\alpha=\beta=1$.  We will
prove the following lemma:
\begin{lemma}
For $L_-' \le L_-$ and $L_+' \le L_+$,
\[
J_{L_-,L_+}(r) = J_{L_-',L_+}(r) + O(r^{L_-' + 2})\quad\hbox{and}\quad
J_{L_-,L_+}(r) = J_{L_-,L_+'}(r) + O(r^{L_+' + 2}).
\]
\end{lemma}
Theorem~\ref{asymtheorem}  is a direct consequence:  for $L_2\ge L_1$,
$J_{L_2}(r)\equiv J_{L_2,L_2}(r) = J_{L_2,L_1}(r) + O(r^{L_1 + 2})
=J_{L_1,L_1}(r) + O(r^{L_1 + 2}) \equiv J_{L_1}(r) + O(r^{L_1 + 2})$.

Now we prove the lemma, considering only the case $L_+' \le L_+$  since
the case $L_-' \le L_-$ is the same by symmetry.  We will need to
consider probabilities of certain collections of configurations.  We write
\begin{equation}\label{marginal-notation}
\Pr ( \eta_{-L_-+1}, \eta_{-L_-+2}, \ldots, \eta_0 ; \eta_1, \eta_2,
\ldots, \eta_N )_{L_-,L_+'}
\end{equation}
for $0\le N\le L_+'$.
This represents the marginal probability of the configuration at sites
$-L_-+1, -L_-+2, \ldots N$; the rest of the sites can take
arbitrary values.  The semicolon indicates the position of the blockage
in the system.

Let $e_R(\eta)$ be the number of the $\eta_1, \eta_2, \ldots, \eta_N$
that take the value 1, {\em i.e.}\ $e_R(\eta)=\sum_{i=1}^N \eta_i$.  For
$r$ small particles to the right of the blockage are rare and can be
treated as excitations; $e_R(\eta)$ can be thought of as the (right)
excitation number in the system.

Our proof proceeds by induction:
Suppose we know {\em all} probabilities of the form\\
$\Pr ( \eta_{-L_-+1}, \eta_{-L_-+2}, \ldots, \eta_0 ; \eta_1,
\ldots, \eta_N )_{L_-,L_+'}$ to a certain accuracy, namely
to order $e_R(\{\eta_1, \ldots, \eta_N\}) + k$ in $r$.  Then (for $N
> 1$) we can (show that we can) compute all probabilities of the form
$\Pr ( \eta_{-L_-+1}, \eta_{-L_-+2}, \ldots, \eta_0 ; \eta_1,
\ldots, \eta_{N-1} )_{L_-,L_+'}$
up to order $e_R(\{\eta_1, \ldots, \eta_{N-1}\}) + k + 1$
in $r$.

This is sufficient to prove our lemma, and in fact more general results:
for any size system (with $L_+' \ge N$) we can compute
$\Pr ( \eta_{-L_-+1}, \eta_{-L_-+2}, \ldots, \eta_0 ; \eta_1,
\ldots, \eta_N )_{L_-,L_+'}$ up to order $e_R(\eta)$ simply by
making use of the fact that excitations are created at rate $r$,
providing the initial step in the induction.
This estimate (and as a result all following estimates) is independent
of $L_+'$ and thus also valid for all $L_+ \ge L_+'$. Since the current
is
\begin{equation}
J(r) = r\Pr (1;0) = r[1 - \Pr (0;0) - \Pr (1;1) - \Pr (0;1)]
\end{equation}
if we can iterate the induction step $j$ times we know the current to
order $2+j$.

So let us return to the induction step.  Consider the transitions that
occur between the different values of $\eta_{-L_-+1}, \ldots
\eta_{N-1}$: we will give estimates on the rates of these transitions in
the stationary state.  (The rate of a transition $\tau\rightarrow\tau'$,
${\rm rate\,} [\tau\rightarrow\tau']$, where $\tau$ and $\tau'$ are sets
of configurations, is
\begin{equation}
{\rm rate\,} [\tau\rightarrow\tau'] = \lim_{\Delta t\rightarrow0}
	(\Delta t)^{-1} \Pr(\tau'\, {\rm at~time\,}t+\Delta t |
	\tau\, {\rm at~time\,}t).
\end{equation}
Thus all (elementary) transitions between individual configurations
occur at rate 0, 1 or $r$.)
Transitions that increase $e_R(\{\eta_1, \ldots,
\eta_{N-1}\})$ involve a jump across the blockage, and thus simply have
rate $r$.  Transitions that keep $e_R(\{\eta_1, \ldots, \eta_{N-1}\})$ fixed
do not involve sites outside of $\{ -L_-+1, -L_-+2, \ldots, N-1 \}$ and
so have rates that are simple integers.  Transitions that decrease
$e_R(\{\eta_1, \ldots, \eta_{N-1}\})$ require that we know something about
site $N$, since the only way to reduce $e_R(\{\eta_1, \ldots, \eta_{N-1}\})$
is for a particle to move from site $N-1$ to site $N$.  So
\begin{eqnarray}\lefteqn{
{\rm rate\,}\left[
        e_R(\{\eta_1, \ldots, \eta_{N-2},1\}) \rightarrow
        e_R(\{\eta_1, \ldots, \eta_{N-2},0\}) \right]}\nonumber\\
&=&     \frac{\Pr ( \eta_{-L_-+1}, \ldots, \eta_0 ; \eta_1,
                \ldots, \eta_{N-2},1,0)_{L_-,L_+'}}
        {\Pr ( \eta_{-L_-+1}, \ldots, \eta_0 ; \eta_1,
                \ldots, \eta_{N-2},1)_{L_-,L_+'}}\label{rate1}\\
&=&     1 - \frac{\Pr ( \eta_{-L_-+1}, \ldots, \eta_0 ; \eta_1,
                \ldots, \eta_{N-2},1,1)_{L_-,L_+'}}
        {\Pr ( \eta_{-L_-+1}, \ldots, \eta_0 ; \eta_1,
                \ldots, \eta_{N-2},1,1)_{L_-,L_+'} +
        \Pr ( \eta_{-L_-+1}, \ldots, \eta_0 ; \eta_1,
                \ldots, \eta_{N-2},1,0)_{L_-,L_+'}}\nonumber
\end{eqnarray}
The numerator of (\ref{rate1}) is $O(e_R(\{\eta_1,\ldots, \eta_{N-1}\})
+ 1)$ and we know it to $O(e_R(\{\eta_1,\ldots, \eta_{N-1}\}) + k + 1)$
by the induction hypothesis.  The denominator of (\ref{rate1}) is
$O(e_R(\{\eta_1,\ldots, \eta_{N-1}\}))$ and we know it to
$O(e_R(\{\eta_1,\ldots,
\eta_{N-1}\}) + k)$ by the induction hypothesis.
Thus the rate in (\ref{rate1}) is $1-O(r)$ and we know it to
$O(e_R(\{\eta_1,\ldots, \eta_{N-1}\}) + k + 1)$.

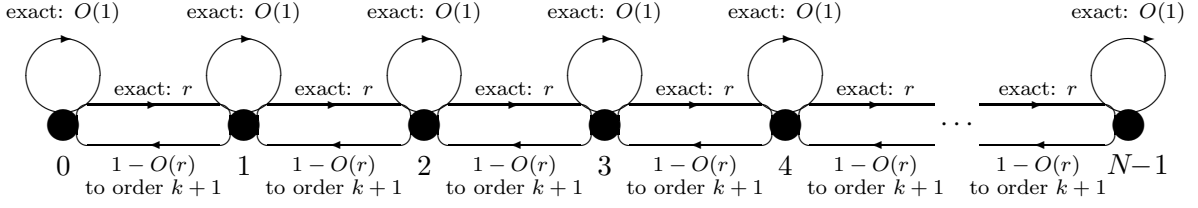
\begin{figure}
\begin{center}
\setlength{\unitlength}{1.2cm}
\begin{picture}(13.7,2.4)(.1,1.35)
\multiput(1,2)(2,0){5}{\circle*{.35}}
\put(10.9,2){\makebox(0,0){$\cdots$}}
\put(12.8,2){\circle*{.35}}
\put(1,1.55){\makebox(0,0){0}}
\put(3,1.55){\makebox(0,0){1}}
\put(5,1.55){\makebox(0,0){2}}
\put(7,1.55){\makebox(0,0){3}}
\put(9,1.55){\makebox(0,0){4}}
\put(12.9,1.55){\makebox(0,0){$N\!\!-\!1$}}
\multiput(2,2.05)(2,0){4}{\oval(1.7,.35)[t]}
\multiput(2,1.95)(2,0){4}{\oval(1.7,.35)[b]}
\put(10,2.05){\oval(1.7,.35)[tl]}
\put(10,1.95){\oval(1.7,.35)[bl]}
\put(11.8,2.05){\oval(1.7,.35)[tr]}
\put(11.8,1.95){\oval(1.7,.35)[br]}
\multiput(10.,2.225)(0,-.45){2}{\line(1,0){.65}}
\multiput(11.8,2.225)(0,-.45){2}{\line(-1,0){.65}}
\multiput(1,2.55)(2,0){5}{\circle{.8}}
\put(12.8,2.55){\circle{.8}}
\multiput(1.09,2.95)(2,0){5}{\vector(1,0){0}}
\put(13.09,2.95){\vector(1,0){0}}
\multiput(2.05,2.225)(2,0){5}{\vector(1,0){0}}
\multiput(1.95,1.775)(2,0){5}{\vector(-1,0){0}}
\put(11.85,2.225){\vector(1,0){0}}
\put(11.75,1.775){\vector(-1,0){0}}
\multiput(1.,3.24)(2,0){5}{\makebox(0,0){\scriptsize exact: $O(1)$}}
\put(12.8,3.24){\makebox(0,0){\scriptsize exact: $O(1)$}}
\multiput(2,2.4)(2,0){5}{\makebox(0,0){\scriptsize exact: $r$}}
\multiput(2,1.55)(2,0){5}{\makebox(0,0){\scriptsize $1-O(r)$}}
\multiput(2,1.3)(2,0){5}{\makebox(0,0){\scriptsize to order $k+1$}}
\put(11.8,2.4){\makebox(0,0){\scriptsize exact: $r$}}
\put(11.8,1.55){\makebox(0,0){\scriptsize $1-O(r)$}}
\put(11.8,1.3){\makebox(0,0){\scriptsize to order $k+1$}}

\end{picture}
\end{center}
\caption{Markov Chain-like representation for the transition rates in
the partial system.  The indices on the ``states'' are the $e_R(\eta)$
which are altered by the ``horizontal'' transitions.  The ``circular''
transitions are those between states with the same $e_R(\eta)$.
}\label{rates}
\end{figure}
Fig.~\ref{rates} thus represents our transitions, where we group
together states with a common $e_R$.  It is clear that it is consistent
to solve for the probabilities in stationary state of the subsystem
represented by fig.~\ref{rates} to $O(e_R(\{\eta_1,\ldots,
\eta_{N-1}\}) + k + 1)$.  This completes the induction step and thus the
proof of the lemma.

The proof of theorem~\ref{asymtheorem2} proceeds in identical fashion.
The only difference is that in (\ref{marginal-notation}) one is limited
to considering $0\le N < L_+'$ instead of $0\le N\le L_+'$, ensuring
that the transition rates for changing the excitation number do not
depend on the boundary conditions.  Thus one can perform the induction
step one time less than for the $\alpha=\beta=1$ case and the boundary
dependence appears one term earlier in the Taylor series.

\renewcommand{\baselinestretch}{1.2}\large\normalsize
\subsection*{Acknowledgments}
We thank F. Alexander, B. Derrida and especially M. Bramson for helpful
discussions.  We also thank the IHES in Bures-sur-Yvette, where part of
this work was done, for its hospitality.

\end{document}